\documentstyle[aps,twocolumn,epsf]{revtex}

\begin{document}
\draft

\title{Creating a low-dimensional quantum gas using dark states in an
inelastic evanescent-wave mirror}

\author{R.J.C. Spreeuw, D. Voigt, B.T. Wolschrijn and H.B. van Linden
van den Heuvell}

\address{Van der Waals-Zeeman Instituut, Universiteit van Amsterdam,\\
Valckenierstraat 65, 1018 XE Amsterdam, the Netherlands\\
e-mail: spreeuw@phys.uva.nl}

\date{\today}
\maketitle

\begin{abstract}

We discuss an experimental scheme to create a low-dimensional gas of
ultracold atoms, based on inelastic bouncing on an evanescent-wave
mirror. Close to the turning point of the mirror, the atoms are
transferred into an optical dipole trap. This scheme can compress the
phase-space density and can ultimately yield an optically-driven atom
laser. An important issue is the suppression of photon scattering due to
``cross-talk'' between the mirror potential and the trapping potential.
We propose that for alkali atoms the photon scattering rate can be
suppressed by several orders of magnitude if the atoms are decoupled
from the evanescent-wave light. We discuss how such dark states can be
achieved by making use of circularly-polarized evanescent waves.

\end{abstract}


\section{Introduction}
\label{sec:intro}

The only route to quantum degeneracy in a dilute atomic gas which has
been experimentally successful so far, is evaporative cooling \cite
{AndEnsMat95,DavMewAnd95,BraSacHul97}. Other routes to quantum
degeneracy, in particular all-optical methods, have been elusive until
now. Nevertheless it is interesting as well as important to keep
exploring alternative methods which do not rely on atomic collisions.
Such systems may be held away from thermal equilibrium and may therefore
constitute a closer matter-wave analogy to the optical laser, as compared
to atom lasers based on Bose-Einstein condensation \cite{MewAndKur97}.
In addition, the physics will be quite different because a different
physical, {\em viz.} optical, interaction would be used to populate the
macroscopic quantum state: the amplification of a coherent matter wave,
while emitting photons.

Several proposals for an optically-driven atom laser have previously
been published. They have in common that a macroscopic quantum state is
populated using an optical Raman transition
\cite{SprPfaJan95,OlsCasDal96}. One problem that has been anticipated
from the beginning, is heating and trap loss caused by reabsorption of
the emitted photons. Therefore later proposals and current experiments
\cite{OvcManGri97,GauHarSch98} have aimed at a reduced dimensionality,
based on optical pumping close to a surface \cite
{DesDal96,PowPfaWil97,PfaMly97,HinBosHug98}. At the same time, there is
also increasing interest in the low-dimensional equivalents of
Bose-Einstein condensation in ultracold gases \cite{SafVasYas98}. Here
we argue that an evanescent-wave mirror is particularly promising for
loading a low-dimensional trap close to a surface.

We extend previous work \cite{SprPfaJan95,OlsCasDal96} so that it can be
applied to the alkali atoms. Being the favorite atoms for laser-cooling,
the application to the alkalis will make this kind of experiments more
easily accessible. In comparison to previous experiments with metastable
rare gas atoms \cite{GauHarSch98}, the alkalis have the advantage that
they do not suffer from Penning ionization. Furthermore, several alkali
species have been cooled to the Bose-Einstein condensation, which makes
them good candidates to create also low-dimensional quantum degeneracy.
The extension to the alkalis is nontrivial because the splitting between
the hyperfine ground states is not large enough to address them
separately with far detuned lasers. The resulting ``cross-talk'' would
lead to large photon scattering rates in the trap, as will be explained
below. We propose to use circularly-polarized evanescent waves and to
trap alkali atoms in ``dark states''. This allows the detuning to be
increased and the photon scattering rate to be reduced by several orders
of magnitude.

Finally, we note that a trap for ultracold atoms close to a surface is
very interesting from the viewpoint of cavity QED. The proximity of a
dielectric surface can change the radiative properties of atoms. In
particular, for circularly-polarized evanescent waves it has been
predicted that the radiation pressure is not parallel to the Poynting
vector \cite{HenCou98}. However, this is beyond the scope of the present
paper.

\section{Generic scheme}
\label{sec:generic}

\subsection{Optical trap loaded by a spontaneous Raman transition}

\begin{figure}
\centerline{\epsfxsize=4cm\epsffile{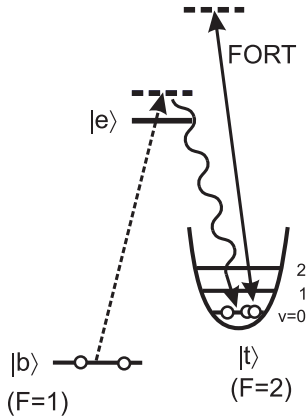}}
\vspace*{0.5cm}
\caption{Three-level scheme, with atomic internal states
$|b\rangle$, $|e\rangle$ and $|t\rangle$, see text. Atoms are
accumulated by means of a spontaneous Raman transition from the unbound
state $|b\rangle$ into the bound levels of a far off-resonant trapping
potential (FORT), operating on atoms in the state $|t\rangle$. Bosonic
enhancement should eventually channel all atoms into the same bound
level $v$.}
\label{fig:generic}
\end{figure}

We start by briefly reviewing the generic idea of loading an optical
atom trap by an optical (Raman) transition. The original proposal
described in Ref.\ \cite{SprPfaJan95} is based on a $\Lambda$-type
configuration of three atomic levels, which we will indicate here by
$|t\rangle $, $|b\rangle $ and $|e\rangle $, as shown in Fig.\
\ref{fig:generic}. The levels $|t\rangle $ and $|b\rangle $ (for
``trapping'' and ``bouncing'' state) are electronic ground (or
metastable) states, $|e\rangle $ is an electronically excited state. An
optical trap is created for atoms in level $|t\rangle $ using the
optical dipole (``light shift'') potential induced by a far
off-resonance laser. Level $|b\rangle $ serves as a reservoir of
ultracold atoms, prepared by laser cooling. The ultracold atoms are
transferred from the reservoir to the trap by a spontaneous Raman
transition $|b\rangle \to |e\rangle \to |t\rangle $.

Our goal is to load a large number of atoms into a single bound state
$|t,v\rangle $ of the trapping potential, where $v$ is the vibrational
quantum number. If the atoms are bosons, the transition probability into
state $|t,v\rangle $ should be enhanced by a factor $1+N_{v}$, where
$N_{v}$ is the occupation of the final state $|t,v\rangle $. If the rate
at which atoms are pumped from $|b\rangle$ to $|t\rangle$ exceeds a
threshold value, the buildup of atoms in $|t,v\rangle$ should rapidly
increase.

The Raman filling process can thus be stimulated by the matter wave in
the trapped final state, leading to matter-wave amplification. The
associated threshold is reached when, for some bound state $|t,v\rangle
$, the unenhanced filling rate exceeds the unavoidable loss rate. The
threshold can be lowered either by decreasing the loss rate or by
increasing the overlap of wavefunctions (``Franck-Condon factor'').

Ideally, the energy separation between states $|t\rangle $ and
$|b\rangle $ should be so large that they can be addressed separately by
different lasers. Examples are alkaline earth atoms or metastable rare
gas atoms. The loading scheme has been applied successfully to load
metastable argon atoms into a far off-resonance lattice
\cite{MulHarBre97} and into a quasi-2D planar matter waveguide
\cite{GauHarSch98}. The two metastable states of Ar* are separated by
42~THz. In this paper we concentrate on $^{87}$Rb atoms, which we use in
our experiments. Here the separation between the two hyperfine ground
states $F=1,2$ is only 6.8~GHz. This requires a modification as will be
discussed in Sec.\ \ref{sec:specific}.

\subsection{The problem of photon reabsorption}

It has been recognized early on that the photon emitted during the Raman
process can be reabsorbed and thus remove another atom from the trap.
This will obviously counteract the gain process and may even render the
threshold unreachable \cite{OlsCasDal96}. This conclusion may be
mitigated in certain situations, such as in highly anisotropic traps
\cite{CasCirLew98}, in small traps with a size of the order of the
optical wavelength \cite{JanWil96} and in the so-called {\em festina
lente} regime \cite{CirLewZol96}.

Our approach is to aim for a low-dimensional geometry, with at least one
strongly confining direction $z$, so that the Lamb-Dicke parameter
$kz_{0} =\sqrt{\omega _{\rm R}/\omega }\ll 1$ in that direction
\cite{VulChiKer98,BouPerKuh99}. Here $k$ is the optical wavevector,
$z_{0}=\sqrt{\hbar /2m\omega }$ is the r.m.s. width of the ground state
of the trap with frequency $\omega $ for an atomic mass $m$, and $\omega
_{\rm R}=\hbar k^{2}/2m$ the recoil frequency. A low-dimensional
geometry should reduce the reabsorption problem because the emitted
photon has a large solid angle available to escape without encountering
trapped atoms. Furthermore, we expect to compress the phase-space
density by loading the low-dimensional optical trap by an
evanescent-wave mirror (``atomic trampoline''), using optical pumping.

\section{Loading a low dimensional trap}
\label{sec:specific}

\subsection{Inelastic evanescent-wave mirror}

We now discuss the specific way in which the generic scheme
discussed above is being implemented in our experiment. Our
implementation is based on an evanescent-wave mirror, using explicitly
the level scheme of $^{87}$Rb atoms. The role of the states $|t\rangle $
and $|b\rangle $ is played by the two hyperfine sublevels of the ground
state $5s\;^{2}S_{1/2}\ (F=1,2)$, which are separated by 6.8~GHz. We
take the lower level, $F=1$ as the ``bouncing state'' $|b\rangle $ and
the upper level, $F=2$, as the ``trapping state'' $|t\rangle $.

\begin{figure}
\centerline{\epsfxsize=3cm\epsffile{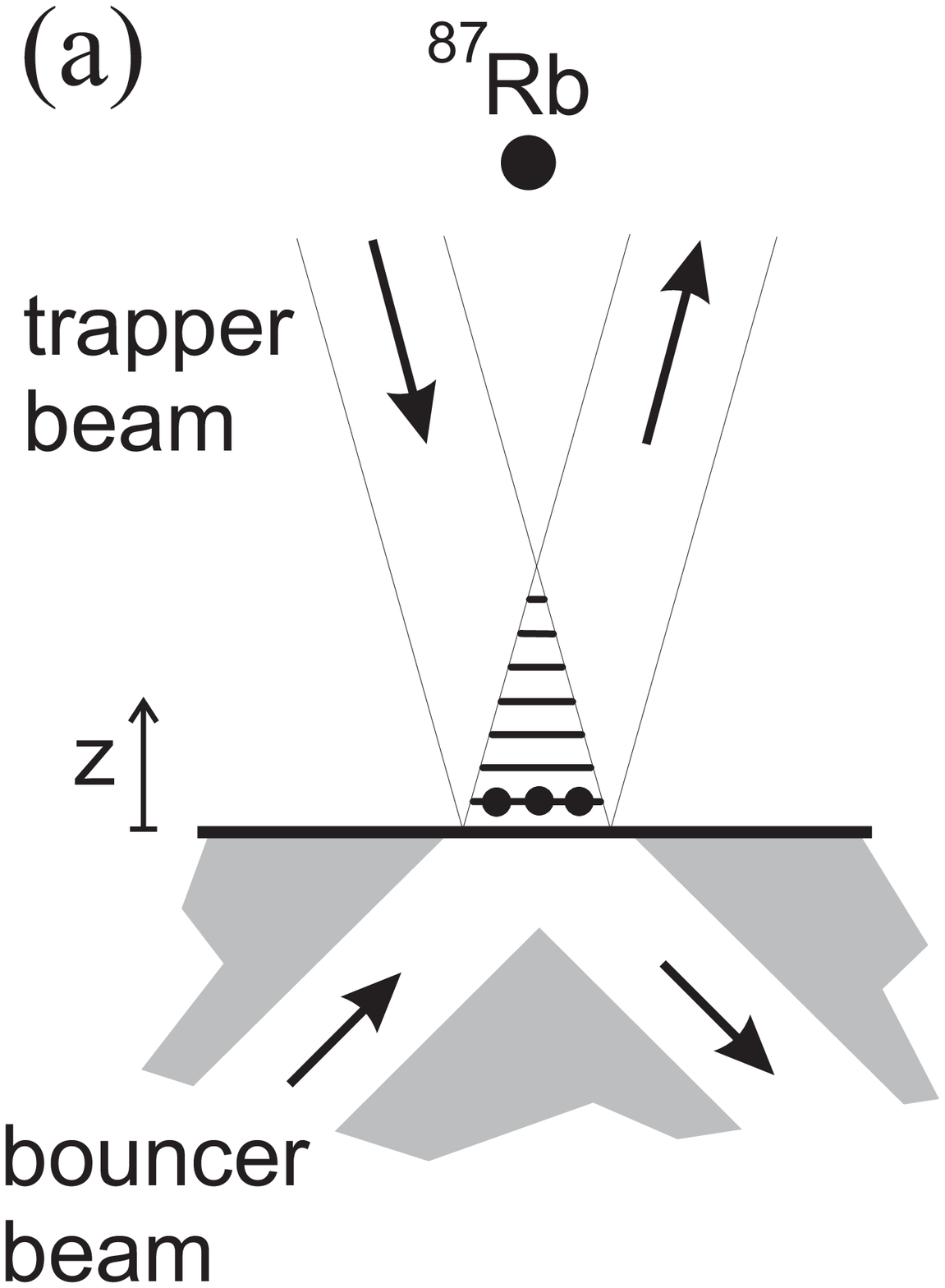}
\hspace*{0.5cm}
\epsfxsize=4.5cm\epsffile{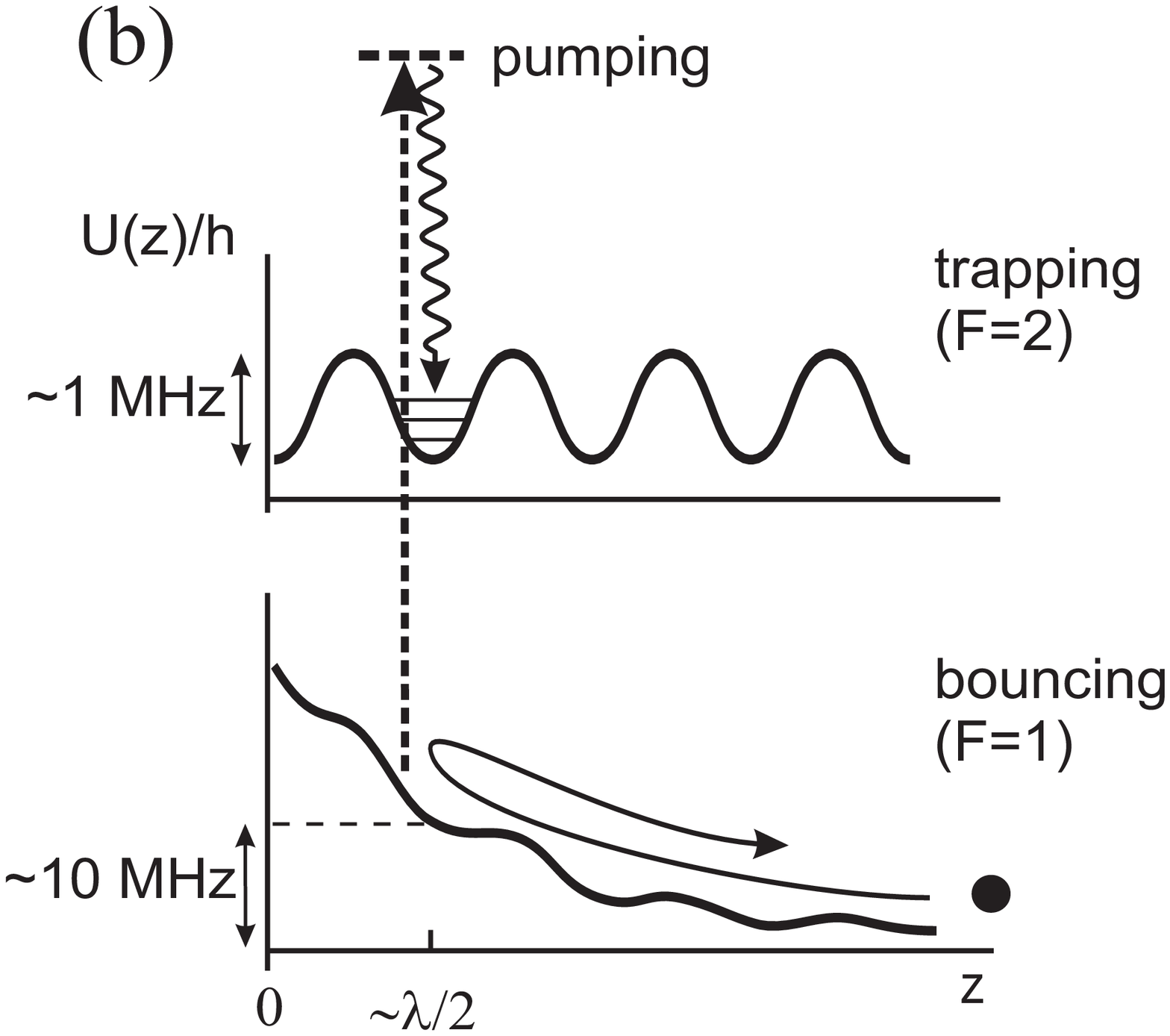}}
\vspace*{0.5cm}
\caption{(a) Geometry of laser beams, incident on a vacuum-dielectric
interface. (b) Corresponding potential curves for the ``bouncing'' and
``trapping'' state ($F=1,2$ for $^{87}$Rb). Ultracold atoms fall towards
the surface, where they are slowed down by the repulsive potential due
to the evanescent ``bouncing'' field. Near the turning point the atoms
undergo a spontaneous Raman transition and become trapped in the optical
potential of a standing ``trapping'' wave. The ripple on the evanescent
wave represents cross-talk from the standing wave, see text. The
tickmark at one half optical wavelength $\lambda /2$ indicates the
typical length scale.} 
\label{fig:specific}
\end{figure}

We consider a configuration of laser beams as sketched in Fig.\
\ref{fig:specific}(a). An evanescent wave is generated by total
internal reflection of a ``bouncer'' beam inside a prism. This bouncer
is blue detuned with respect to a transition starting from the $F=1$
ground state, with detuning $\delta _{1}$. A second laser beam, the
``trapper'' beam, is incident on the prism surface from the vacuum side
and is partially reflected from the surface. The reflected wave
interferes with the incident wave to produce a set of planar fringes,
parallel to the prism surface. Note that even with 4\% reflectivity of
an uncoated glass surface the fringe visibility will be 0.38.

The trapper beam can be either red or blue detuned, the former having
the advantage that it automatically provides transverse confinement. In
Fig.~\ref{fig:specific} we sketched the situation for blue detuning,
confining the atoms vertically in the intensity minima, but allowing
them to move freely in the transverse direction. We assume that the loss
rate due to moving out of the beam is slow compared to other loss rates,
such as that due to photon scattering. Alternatively one can obtain
transverse confinement by using multiple trapper beams from different
directions, which interfere to yield a lattice potential. Similarly, one
can create an optical lattice using multiple bouncer beams, see e.g.
Fig.\ \ref{fig:TEcross}.

Ultracold atoms, in the bouncing state $F=1$, are dropped onto the prism
and are slowed down by the repulsive light-shift potential induced by
the bouncer beam, see Fig.\ \ref{fig:specific}(b). If the potential is
strong enough, the atoms turn around before they hit the prism and
bounce back up. This is called an ``evanescent-wave mirror'', or
``atomic trampoline'' and has been demonstrated by several groups
\cite{KasWeiChu90,HelRolGol92,AmiSteBou93}.

We are here interested in interrupting the bouncing atoms halfway during
the bounce, near the classical turning point. The interruption can occur
when the atom scatters an evanescent-wave photon and makes a Raman
transition to the other hyperfine ground state, $F=2$. This Raman
transition yields a sudden change of the optical potential, because for
an atom in $F=2$ the detuning is larger by approximately the ground
state hyperfine splitting. This mechanism has been used for
evanescent-wave cooling \cite{OvcManGri97}. In our case, we tailor
the potentials so that the bouncer potential dominates for $F=1$ and the
trapper for $F=2$. The atom is thus slowed down by the bouncer and then
transferred into the trapping potential.

As long as the probability for undergoing a Raman transition during
the bounce is not too large, $p\lesssim 1-e^{-2}$, the transition
will take place predominantly near the turning point, for two reasons.
Firstly, the atoms spend a relatively long time near the turning point.
Secondly, the intensity of the optical pump (the evanescent wave) is
highest in the turning point. The probability that the atoms end up in
the lowest bound state of the trapping potential has been estimated to
be on the order of $10-20\%$, albeit for somewhat different geometries
\cite{DesDal96,PowPfaWil97}. The resulting compression of a
three-dimensional cloud into two dimensions is in fact {\em
dissipative}, i.e.\ it can increase the phase-space density.

\subsection{Phase space compression}

\begin{figure}
\centerline{\epsfxsize=8cm\epsffile{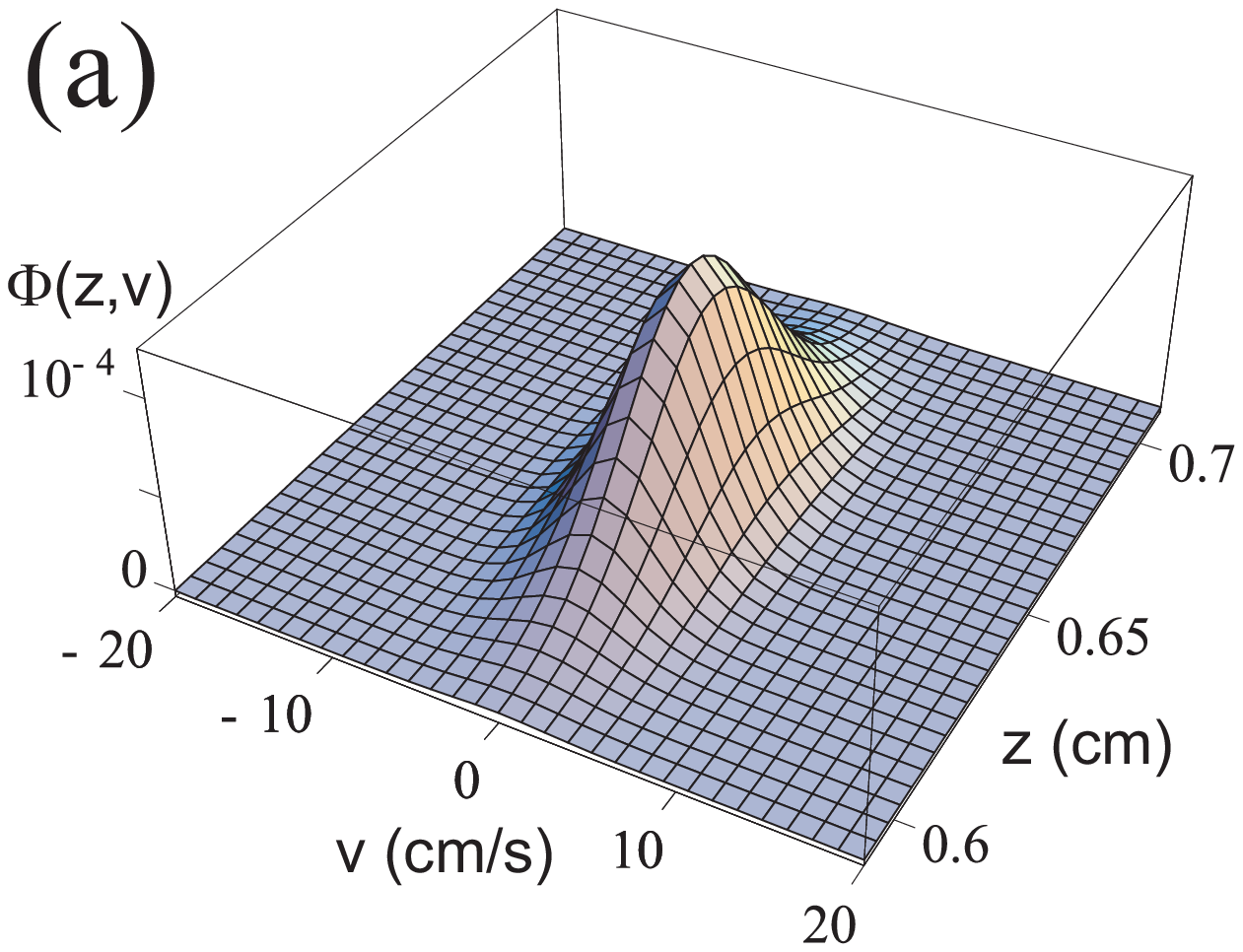}}
\centerline{\epsfxsize=8cm\epsffile{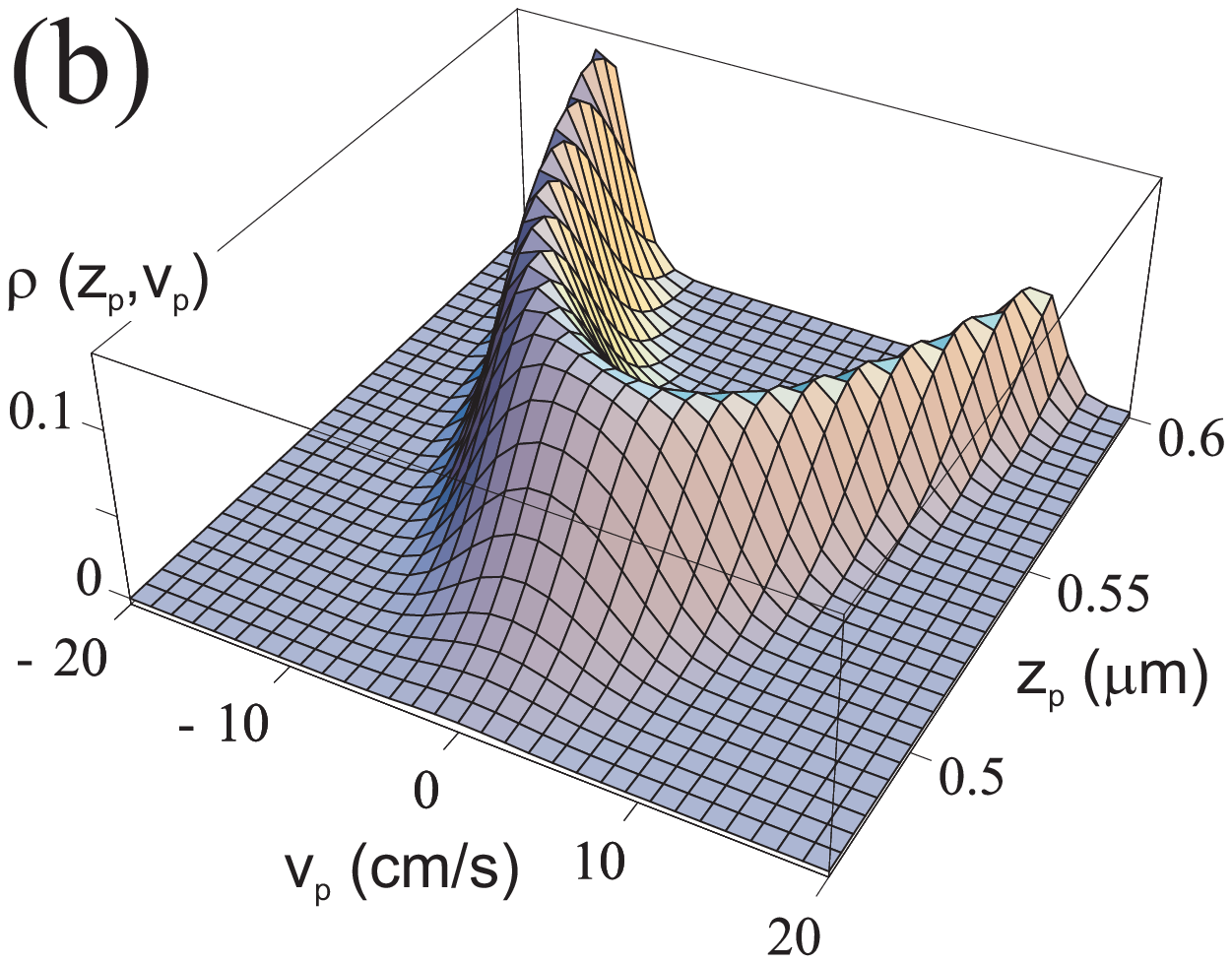}}
\vspace*{0.3cm}
\caption{The phase-space compression due to inelastic bouncing on an
evanescent-wave mirror, based on a classical trajectory simulation. (a)
Initial one-dimensional phase-space distribution of a single atom. (b)
Distribution of phase-space coordinates where the bounce is interrupted 
due to a spontaneous Raman transition. Note that the spatial scale
changes from cm to $\mu$m and that the peak phase-space density along
the line $v=0$ increases by a factor $\sim 10^3$.} 
\label{fig:compressor} 
\end{figure}

As an illustration we give the result of a classical trajectory
simulation. We start from the dimensionless phase-space distribution 
$\Phi (z,v)$ for the vertical motion of a single atom cooled in optical
molasses, shown in Fig.\ \ref{fig:compressor}(a). Note that $v$ denotes
here the vertical velocity component and that we drop the subscript
$z$ in $v_z$ throughout this paper. The phase-space density has been
made dimensionless by dividing it by the phase-space density of quantum
states. The latter is given by $m/h$ (states per unit area in the
$(z,v)$ space), where $h$ is Planck's constant.  The distribution $\Phi
(z,v)$ can be interpreted as the probability that the atom is in an
arbitrary quantum state localized around $(z,v)$.

The atom, described by the classical distribution $\Phi (z,v)$, is
assumed to enter the evanescent wave at a velocity $v_{i}=p_{i}/m$,
determined by its velocity in the molasses $v$ and the height $z$ from
which it falls. Inside the evanescent wave the atom moves as a point
particle along a phase-space trajectory $(z(t),v(t))$, governed by the
evanescent-wave potential.

Assuming that the saturation parameter is small, the potential is given
by $U(z)=U_{0}\exp (-2\kappa z)$, where  $\kappa=k_L\sqrt{n^2 \sin^2
\theta_i -1}$ is the decay constant of the evanescent wave, with $k_L$
the free-space laser wavevector, $\theta_i$ the angle of incidence and
$n$ the index of refraction. Similarly, the photon scattering rate is
given by $\Gamma'(z)=\Gamma'_0\exp (-2\kappa z)$, with
$\Gamma'_0/U_0=\Gamma/\hbar\delta$, where $\Gamma$ is the natural
linewidth and $\delta$ is the laser detuning. Finally, the Raman
transition rate is given by $R(z)=R_0\exp (-2\kappa z)$, with
$R_0=b\Gamma'_0$, where $b$ is the branching ratio, i.e. the probability
that photon scattering leads to a Raman transition. The Raman rate gives
the local probability per unit time that the trajectory is interrupted.

Due to the stochastic nature of the spontaneous Raman transition, we
obtain a probability distribution over pumping coordinates where the
trajectory through phase space is interrupted due to the transition,
$\rho (z_p,v_p)$. Not surprisingly, we see in Fig.\
\ref{fig:compressor}(b) that this distribution has the shape of a
``mountain ridge'' following the phase-space trajectory.

Our goal is to load the pumped atoms into a bound state of a trap near
the surface. Therefore the number of interest is the peak value of $\rho
(z_p,0)$, which occurs for a value of $z$ near the turning point. In
Fig.\ \ref{fig:compressor}(b) we see that the peak value of $\rho
(z_p,0)$ is about $1000$ times higher than the initial peak value of
$\Phi(z,0)$ in optical molasses (Fig.\ \ref{fig:compressor}(a)). The
peak value of 0.11 can be interpreted as the trapping probability in the
ground state of the trap that collects the atoms. This value is quite
comparable to previous calculations by different methods
\cite{DesDal96,PowPfaWil97}.

The position of the turning point should be adjusted to coincide with
the center of the trap, for example by adjusting $U_0$ or $\kappa$. The
trapping probability can be maximized by changing the value of $\kappa $
and/or the ratio $U_{0}/R_{0}$, in such a way that $U_{0}/R_{0}=m
v_i/2\kappa$. This corresponds to a situation where the probability for
reaching the turning point without being optically pumped is $e^{-1}$.
If the pumping rate is very high, too many atoms are pumped before they
reach the turning point. If the pumping rate is very low, too many atoms
bounce without being pumped at all. If the optical pumping is done by
the same laser that induces the bouncing potential, we have
$U_0/R_0=\hbar \delta /\Gamma b$, so that we obtain an optimum value for
the detuning:\ $\delta /\Gamma =b p_{i}/2\hbar \kappa $. Experimentally
it may be advantageous to use separate lasers for the mirror potential
and for pumping so that this restriction on the detuning does not apply.

Obviously we should be somewhat careful in assigning quantitative
meaning to the result of our classical simulation. In particular we
should verify that the distribution $\rho (z_p,0)$ is broad on the
characteristic length scale of the atomic wavefunction near the turning
point. The latter is determined by the slope of the bouncing potential
near the turning point and is given by $\sim \kappa ^{-1}(\hbar \kappa
/p_{i})^{2/3}.$ For the same parameters as used in Fig.\
\ref{fig:compressor}(b) this characteristic width is $\sim 22$~nm,
indeed smaller than the width of $\rho (z_p,0)$, which is $\sim 50$~nm.

\section{Photon scattering}

\subsection{Metastables versus alkalis}

The level scheme used in the proposal of Ref.~\cite{SprPfaJan95} was
inspired by metastable rare gas or alkaline earth atoms. In those cases
two (meta)stable states can usually be found with a large energy
separation. This makes it relatively straightforward to separate the
bouncing and trapping processes. We extend those ideas here, applying
them to the typical level scheme of the alkalis. In this case the
separation between two stable states is limited to the ground state
hyperfine splitting.

We therefore turn to the issue of photon scattering by atoms after they
have been transferred into the trap. More specifically, our main concern
is scattering of {\em bouncer} light. The rate of scattering light from
the trapping laser can in principle be made negligibly small by choosing
a large enough detuning. This can be done because the trapping potential
can be much shallower than the bouncing potential and therefore need not
be $F$-state specific. For example, if the atoms are dropped from 6~mm
above the prism, their kinetic energy will be 0.6~mK, corresponding to a
minimum bouncing potential of 12~MHz. For the trapping potential, on the
other hand, a depth of less than 1~MHz should be sufficient, since most
of the external energy of the atom has been used for climbing the
bouncing potential. For the bouncing state, $F=1$, the trapping potential
then appears as a small ripple superimposed on the bouncing potential.

The scattering of bouncer light is more difficult to avoid. Ideally, the
interaction of the atoms with the bouncer should vanish completely as
soon as they are transferred into the $F=2$ state. In reality, however,
the bouncer connects both ground states, $F=1$ and $F=2$ to the excited
state through a dipole-allowed transition. We can approach the ideal
situation by a proper choice of the bouncer detuning. For the simplified
three-level scheme of Fig.\ \ref{fig:generic}, a limitation is
imposed by the ground state hyperfine splitting $\delta _{\rm GHF}=2\pi
\times 6.8$~GHz. A good distinction between the $F=1$ and $F=2$ states
is only obtained if the bouncer detuning is small, $\delta_1 \ll
\delta_{\rm GHF}$. However, a very small detuning is undesirable because
it leads to an increased photon scattering rate and thus heating during
the bounce.

The number of photons scattered during the bounce is approximately given
by $\bar{n}_{\rm sc}\simeq (\Gamma /\delta _{1})p_{i}/\hbar \kappa ,$
where $p_{i}\simeq 60\hbar k_{L}$ is the momentum of a Rb atom falling
from a height of about 6 mm and $\kappa \simeq 0.15$ $k_{L}$ (for an
angle of incidence $\theta_i =\theta_c +0.01$). If we operate in the
regime $b\bar{n}_{\rm sc}\simeq 2$ (i.e. until the turning point we have
$b\bar{n}_{\rm sc}\simeq 1$) and set $b=0.5$, this requires a detuning
$\delta _{1}\simeq 100\ \Gamma \simeq 2\pi \times 0.6$~GHz. After the
atom has been transferred into the trapping potential for $F=2$, the
detuning of the bouncer will be $\delta_2=\delta_1+\delta_{\rm
GHF}\simeq 2\pi \times 7.4$~GHz $\simeq 1200\ \Gamma $. The trapped atoms
will then scatter bouncer light at an unacceptably high rate of
typically $5\times 10^{3}$~s$^{-1}$.

\subsection{Dark states}

The limitation imposed by the hyperfine splitting, $\delta _{1}\ll
\delta _{\rm GHF}$, can be overcome by making use of dark states. This
requires a more detailed look at the Zeeman sublevels of the hyperfine
ground states. We consider the state $|F=m_{F}=2\rangle $ and tune the
bouncer laser to the D1 resonance line (795~nm, $5s\
^{2}S_{1/2}\rightarrow 5p\ ^{2}P_{1/2}$). If this light is $\sigma ^{+}$
polarized, the selection rules require an excited state
$|F'=m_{F'}=3\rangle $, which is not available in the $5p\ ^{2}P_{1/2}$
manifold and so $|F=m_{F}=2\rangle $ is a dark state with respect to the
entire D1 line.

The state selectivity of the interaction with bouncer light now no
longer depends on the detuning, but rather on a selection rule.
Therefore the bouncer detuning can now be chosen large compared to
$\delta _{\rm GHF}$. The new limitation on the detuning is the fine
structure splitting of the D-lines, 7.2~THz for Rb. This reduces the
photon scattering rate by 3 orders of magnitude. Note that the heavier
alkalis are more favorable in this respect because of the larger fine
structure splitting. The price to be paid is the restriction to two
specific Zeeman sublevels $|F=\pm m_{F}= 2\rangle $ and the need for a
circularly-polarized evanescent wave.

\section{Circularly-polarized evanescent waves}

In this section we briefly describe two methods for the generation of
evanescent waves with circular polarization, using either a single
bouncer beam or a combination of two. We also calculate the resulting
photon scattering rates. There are several other ways to generate
circularly-polarized evanescent waves using multiple beams. The two
methods described here serve as examples. 

\subsection{Single beam}

A circularly-polarized evanescent wave can be obtained using a single
incident laser beam if it has the proper elliptical polarization, i.e.
the proper superposition of $s$ and $p$ polarization. The $s$, or TE,
mode yields an evanescent electric field parallel to the surface and
perpendicular to the plane of incidence. The evanescent field of the
$p$, or TM, mode is elliptically polarized in the plane of incidence,
with the long axis of the ellipse along the surface normal.

It is straightforward to calculate the input polarization that yields
circular polarization in the evanescent wave. One finds that the
required ellipticity of the input polarization is the inverse of the
refractive index, $n^{-1}$. Here the ellipticity is defined as the ratio
of the minor and major axes of the ellipse traced out by the electric
field vector. The required orientation $\varphi $ of the ellipse depends
on the angle of incidence, $\tan \varphi =-\sqrt{n^{2}\sin ^{2}\theta
_{i}-1}/\cos \theta _{i}$. Close to the critical angle, $\varphi\approx
0$, and the ellipse has its major axis perpendicular to the plane of
incidence.
 
Following this prescription, the resulting evanescent wave will be
circularly polarized, with the plane of polarization perpendicular to
the surface. However, the plane of polarization is not perpendicular to
the in-plane component of the $k$-vector. Here the evanescent wave
differs from a propagating wave, which has its plane of polarization
always perpendicular to the $k$-vector (and Poynting vector). For the
evanescent wave the plane of circular polarization is also perpendicular
to the Poynting vector. However, the Poynting vector is not parallel to
the in-plane $k$-vector, but tilted sideways by an angle $\pm\chi$ given
by $\tan \chi =\sqrt{n^{2}\sin ^{2}\theta _{i}-1}= \kappa \lambda
_{0}/2\pi$, with $\lambda_0$ the vacuum wavelength of the light. Close
to the critical angle, $\chi\approx 0$ and the plane of polarization
becomes perpendicular to the in-plane wave vector, as for propagating
waves.

We can estimate the photon scattering rate of an atom in the dark state
$|F=m_{F}=2\rangle $, residing in the circularly polarized evanescent
wave of the bouncer beam. Ideally, this scattering rate is due only to
off-resonant excitation to the $5p\ ^{2}P_{3/2}$ manifold (D2 line;
780~nm.) Choosing the bouncer detuning at 100~GHz (with respect to the
D1 line) yields a scattering rate $\Gamma'_{D2}=3.5$~s$^{-1}$. In
practice there will also be scattering due to polarization impurity. For
example, assuming this impurity to be 10$^{-3}$, we obtain a scattering
rate of $\Gamma'_{D1,\sigma ^{-}}=10.6~$s$^{-1}.$

\subsection{Two crossing $s$-waves}

\begin{figure}
\centerline{\epsfxsize=7cm\epsffile{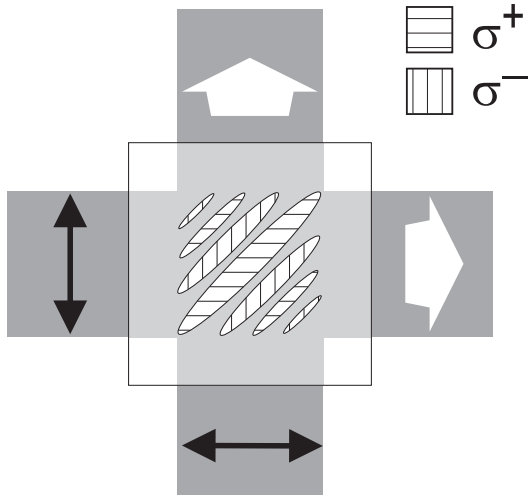}}
\vspace*{0.5cm}
\caption{Generating circularly-polarized evanescent waves by crossing
two TE-polarized waves at a right angle (looking down at the prism
surface). The polarizations and in-plane wave vector components yield a
fringe pattern of alternating lines of opposite circular polarization. The
total intensity is constant across the pattern, since the two TE
polarizations are orthogonal.}
\label{fig:TEcross}
\end{figure}

Alternatively, evanescent waves of circular polarization can also be
produced using two (or more) bouncer beams. If we cross two TE-polarized
evanescent waves at 90${^{\circ }}$, we will produce a polarization
gradient as sketched in Fig.\ \ref{fig:TEcross}. Lines of circular
polarization are now produced with the plane of polarization {\em
parallel} to the surface. Lines of opposite circular polarizations
alternate, with a distance of approximately $\lambda_0 /2\sqrt{2}$
between neighbouring $\sigma ^{+}$ and $\sigma ^{-}$ lines. 

This configuration offers interesting opportunities. The light field can
be decomposed into two interleaved standing wave patterns, one for
$\sigma ^{+}$ and one for $\sigma ^{-}$ polarization. An atom in the
state $|F=m_{F}=2\rangle $ is dark with respect to the $\sigma ^{+}$
standing wave only. However it does interact with the $\sigma ^{-}$
standing wave and therefore can be trapped in its nodes. The bouncer
light will thus play a double role. First it slows the atoms on their
way down to the surface. Then, after the atoms have been optically
pumped, the bouncer light will transversely confine the atoms. We thus
expect a one-dimensional lattice of atomic quantum wires with
alternating spin states, very much like a surface version of previously
demonstrated optical lattices \cite{JesGerLet92,VerLouSal92,JesDeu96}.
The transverse lattice structure may also allow the use of transverse
Sisyphus cooling or Raman sideband cooling
\cite{VulChiKer98,HamHayKlo98}. 

It is not strictly necessary to cross the evanescent waves at a right
angle. It does have the advantage that the total intensity is constant
across the polarization pattern. The same could also be achieved by
using counterpropagating evanescent waves with orthogonal polarizations.
For any other angle, the intensity varies spatially so that the atoms
bounce on a corrugated optical potential. However, even with a uniform
intensity, most atoms will see a corrugated potential. The potential
depends on the local polarization and on the atom's magnetic sublevel
through the Clebsch-Gordan coefficients. Only for the state
$|F=1,m_{F}=0\rangle $ the dipole potential is independent of the
polarization. One could of course prepare the falling atoms in
$|F=1,m_{F}=0\rangle $ using optical pumping. The local circular
polarization $\sigma ^{\pm}$ will tend to pump the atom into the local
dark state $|F=\pm m_{F}=2\rangle $. However the optical pumping
transition has a branching ratio of only 1/6 (using a dedicated resonant
pumping beam). By contrast, for an atom starting in $|F=m_{F}=1\rangle
$, the branching ratio is 1/2. Therefore starting in
$|F=1,m_{F}=0\rangle $ is conceptually simple, but probably not optimal.

A disadvantage of creating circularly polarized evanescent waves on a
lattice is that an additional source of photon scattering appears. We
approximate the transverse potential near the minimum as a harmonic
oscillator. Choosing again the bouncer detuning at 100~GHz, the harmonic
oscillator frequency will be about $\omega=2\pi\times 480$~kHz. An atom
in state $|2,2\rangle $, in the ground state of the harmonic oscillator
associated with the $\sigma ^{-}$ node has a gaussian wavefunction with
wings extending into the region with $\sigma ^{-}$ light. The resulting
scattering rate can be estimated as $\Gamma'_{\rm HO}\approx
\frac{1}{4}\omega \Gamma /(\delta_1+\delta _{\rm GHF})\approx 52$
~s$^{-1}$, where the bouncer detuning was again chosen at 100~GHz. The
scattering rate can be further suppressed to $\Gamma'_{\rm HO}\approx
18$~s$^{-1}$ by raising the bouncer detuning to 300~GHz. For an even
larger detuning the off-resonant scattering by the D2 line,
$\Gamma'_{D2}$, starts to dominate.

\subsection{Feasibility}

We should point out that our two examples to produce
circularly-polarized evanescent waves are not meant to be exhaustive.
Several other methods can be devised, some being more experimentally
challenging than others.

For the single-beam method the incident beam must be prepared with the
correct ellipticity as well as the correct orientation. It will probably
be difficult to measure the polarization of the evanescent wave
directly. One should therefore prepare the incident polarization using
well-calibrated optical retarders and using calculated initial settings.
The fine-tuning could then be done, e.g. by optimizing the lifetime of
the trapped atoms in the dark state. 

For the two-beam method of Fig.\ \ref{fig:TEcross} we have assumed for
simplicity that the two interfering evanescent waves have the same decay
length (i.e. the same angle of incidence) and the same amplitude. Equal
decay lengths for the two waves can be enforced by making use of a
dielectric waveguide \cite{SeiKaiAsp94}. Alternatively, one may
deliberately give the two beams a slightly unequal decay length, and at
the same time give the wave with the shorter decay length a larger
amplitude. In this case there will always be one particular height above
the surface where the two beams have equal amplitude, as required for
interfering to circular polarization. Therefore this procedure would
make the circular polarization somewhat self-adjusting. The height where
circular polarization occurs is tunable by changing the relative
intensity of the two beams. 

Obviously, the final word on the feasibility can only be given
experimentally. In our experiment we are presently pursuing a variation
on Fig.\ \ref{fig:TEcross}, including the just mentioned self-adjusting
properties.

\section{Conclusion}

In conclusion, we have shown that inelastic bouncing on an
evanescent-wave mirror (``atomic trampoline'') is a promising method for
achieving high phase-space density in low-dimensional optical traps.
The phase space compression is achieved by means of a spontaneous Raman
transition, which is highly spatially selective for atoms near the turning
point of the evanescent-wave mirror potential. 

We have extended previous work based on the level schemes of metastable
rare gas atoms for application to alkali atoms. This requires
suppression of the high photon scattering rate, resulting from the
relatively small ground hyperfine splitting in the alkalis. We have
shown how the photon scattering rate can be reduced by several orders of
magnitude, by trapping the atoms in dark states. This requires the use
of circularly-polarized evanescent waves, which can be generated by
several methods as discussed. If built up from multiple beams, the
evanescent field may play a double role, generating a bouncing as well
as a trapping potential. This could lead to an array of quantum wires
for atoms.

\section{Acknowledgments}

This work is part of the research program of the ``Stichting voor
Fundamenteel Onderzoek van de Materie'' (Foundation for the
Fundamental Research on Matter) and was made possible by financial
support from the ``Nederlandse Organisatie voor Wetenschappelijk
Onderzoek'' (Netherlands Organization for the Advancement of
Research). The research of R.S. has been made possible by
financial support from the Royal Netherlands Academy of Arts and
Sciences.

\bibliographystyle{prsty}


\end{document}